\documentclass[a4paper]{spie}

\usepackage[english]{babel}
\usepackage[latin2]{inputenc}
\usepackage[T1]{fontenc}

\usepackage[usenames]{color}

\usepackage{ae,aecompl}
\usepackage{enumerate}
\usepackage{epsfig}
\usepackage{graphics}
\usepackage{graphicx}
\usepackage{amsmath}
\usepackage{amssymb}
\usepackage{pifont}
\usepackage{wasysym}
\usepackage{color}

\title{The limit order book on different time scales} 

\author{Zolt\'an Eisler\supit{a}, J\'anos Kert\'esz\supit{a,b} and Fabrizio Lillo\supit{c,d}
\skiplinehalf
\supit{a}Department of Theoretical Physics, Budapest University of Technology and Economics, Budapest, Hungary; \\
\supit{b}Laboratory of Computational Engineering, Helsinki University of Technology, Espoo, Finland; \\
\supit{c}Dipartimento di Fisica e Tecnologie Relative, Universit\`a di Palermo, Palermo, Italy; \\
\supit{d}Santa Fe Institute, Santa Fe, NM 87501, USA
}

\begin{document}

\authorinfo{Send correspondence to Z.E.: E-mail: eisler@maxwell.phy.bme.hu, Telephone: +36 1 463 2691\\}

\maketitle 

\begin{abstract}
Financial markets can be described on several time scales. We use data from the limit order book of the London Stock Exchange (LSE) to compare how the fluctuation dominated microstructure crosses over to a more systematic global behavior.
\end{abstract}


\maketitle

\section{Introduction}

 


The dynamics of financial markets presents a long standing puzzle that is yet to be understood. Economic systems do not lend themselves to simple explanations, similarly to other complex systems that "evolved to survive, and not for scientists to understand" \cite{alon.simplicity}. The most ambitious goal would be to provide a solid microscopic theory of market participants, their interactions, supply and demand by integrating elements from economics and other areas. Based on this theory one could build from bottom up and see how the many rapidly fluctuating, interacting individuals create the long-term behavior, and finally contribute to macroeconomics. Such a unified theory seems currently out of reach, but there has been much progress on different levels. Behavioral finance and experimental economics \cite{barberis.thaler} have clarified many aspects of how individuals make their decisions. At the other end of the spectrum, regarding the market as a whole, econometrics and more recently econophysics have uncovered several "stylized facts" of prices and trading activity \cite{campbell.lo.mackinlay, bouchaud.book}. By analyzing the structure of limit order books recent studies hope to provide some of the missing links between these two levels \cite{biais.lob, maslov.pa2001, zovko.farmer, farmer.theory, bouchaud.statistical, mike.empirical}. 

At present every level of description has a characteristic time scale on which its models are valid. The typical price change on these durations acts as another important scale. In this paper we will present a comparison of the phenomenology of stock market data as we go from monthly to tick-by-tick resolution. We use limit order book data which contains the maximal amount information about the state of the market that is available to traders. Our aim is to show how much of this microstructure is important at the different time scales. We present qualitative evidence that mainly supports recent theories, but also points out some of their limitations.

The paper is organized as follows. Section \ref{sec:data} presents our dataset. Then Sections \ref{sec:monthly}, \ref{sec:daily} and \ref{sec:intraday} describe the character of the limit order book on monthly, daily and intraday time scales, respectively. Finally, Section \ref{sec:conclusions} concludes. For the readers' convenience a comparison of the typical time and price scales is shown in Tables \ref{tab:time} and \ref{tab:price}. 

\section{Order book data}
\label{sec:data}

This study is based on the trading data of the electronic market (SETS) of London Stock Exchange (LSE) during the year $2002$. The trading day of LSE is divided into three periods. The continuous auction runs during 8:00--16:30, in this paper we will only consider these intervals. The two further periods, the opening and a closing auction during 7:50--8:00 and 16:30--16:35 were discarded. Throughout the paper time will be measured as continuous trading time, starting from 8:00:00 on January $2$, $2002$\footnote{Every trading day lasts exactly $31501$ seconds, except for December $24$, $2002$ and December $31$, $2002$, which are $17101$ seconds long.}.

As units of price we will use ticks, which correspond to $1$ penny in the case of GSK. We intentionally use linear price and not its logarithm. We do so in order to keep tick sizes constant, the minimal change in the log-price would depend on its value. We investigated several liquid stocks, but here we will only present one: GlaxoSmithKline (GSK). The results are qualitatively similar for all stocks where the tick/price ratio is small enough.

The order book is a deterministic mechanism to organize the auction market based on buy and sell limit orders. These are offers to buy or sell a fixed volume of a stock at a given (limit) price. Limit orders with more favorable prices take priority, and for the same price orders that were placed earlier are executed first. The highest price to buy is usually called the best bid price $b_t$, and the lowest price to sell the best ask price $a_t$. The difference of the two is called the bid-ask spread: $s_t = a_t-b_t$. Besides limit orders there also exist so called market orders. These are also characterized by a fixed volume, but they have no preassigned price. Instead, they are requests to buy/sell immediately at the best price available. It is also possible to have limit orders placed with prices that cross the spread (crossing orders). These can be immediately paired with orders from the other side of the book. They can be executed completely if there was enough volume available for trade. Otherwise their remaining part stays in the book as 
a new limit order.

\begin{table}[ptb]
\centering
\begin{tabular}{|c|c|}
\hline
process/event & time scale (sec) \\ \hline
trading day & $3.15 \times 10^4$ \\
crossover to diffusivity & $3\times 10^2 - 10^3$ \\
relaxation of liquidity fluctuations & $10^2-10^3$ \\
intertrade time & $10$ \\
\hline
\end{tabular}
\vskip5mm
\caption{Typical time scales for the stock GSK in the year $2002$. Intertrade times were calculated as a mean for the dataset, over the $3000-28000$'th seconds of every trading day. The time scale for the relaxation of liquidity fluctuations was taken from Ref. \cite{ponzi.liquidity}. The crossover time to diffusivity was taken from Ref. \cite{eisler.orderbook}.}
\label{tab:time}
\end{table}
\begin{table}[ptb]
\vskip10mm
\centering
\begin{tabular}{|c|c|}
\hline
process/quantity & price scale (ticks) \\ \hline
midquote price & $1.4 \times 10^3$ \\
median-to-median width of book & $68$ \\
$1$-month absolute return & $1.4 \times 10^2$ \\
$1$-day absolute return & $23$ \\
distance of "stripes" & $5$ \\
$10$-min absolute return & $3$ \\
bid-ask spread & $1.9$ \\
$1$-min absolute return & $0.9$ \\
\hline
\end{tabular}
\vskip5mm
\caption{Typical time scales for the stock GSK in the year $2002$. The distance of "stripes" was determined from Fig. \ref{fig:pageshot2_long3}. All other quantities represent means for the dataset, over the $3000-28000$'th seconds of every trading day.}
\label{tab:price}
\end{table}

\section{Monthly and longer time scales}
\label{sec:monthly}

On monthly or longer time horizons stock prices are often modeled as an ordinary
diffusion process with a drift \cite{bouchaud.book, black.scholes}. Although this is an extreme
simplification of the actual behavior, it is true that the autocorrelations of
volatility and the fat tails of the return distribution are effectively
diminished on such a long time scale, and so they need not be taken into account. 
The difference between bid and ask, opening and closing prices is also insignificant: the typical monthly
absolute return is two orders of magnitude larger than the spread and about one order of magnitude larger than the difference between the opening and closing prices. 

\begin{figure*}[ptb]
\centerline{\includegraphics[width=300pt,angle=-90]{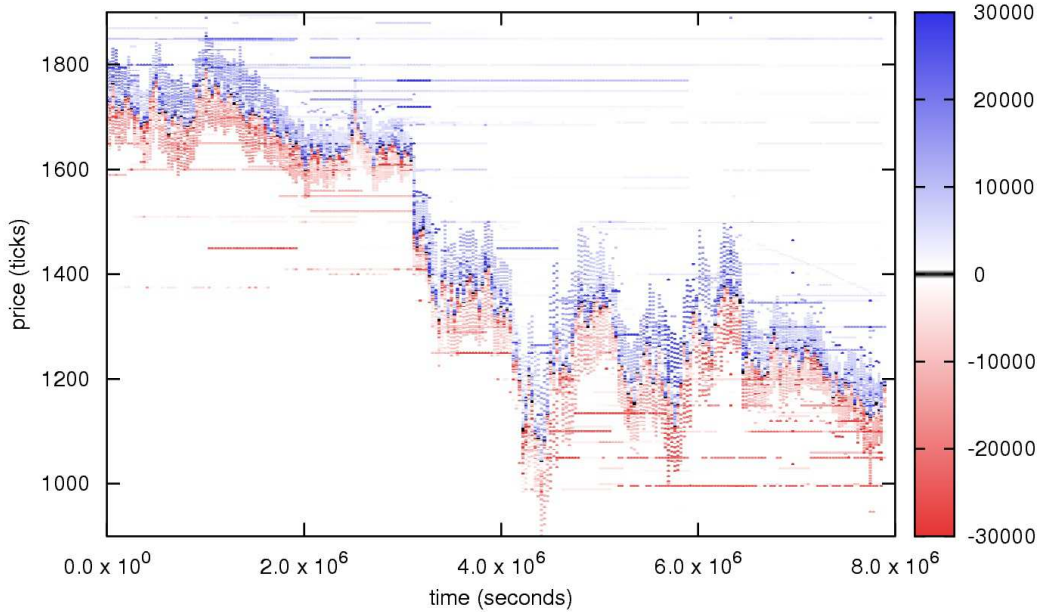}}
\caption{The order book of the stock GSK during the year $2002$. For all $252$ trading days during the year we made a snapshot of the order book at the $15000$'th second of the day. The lack of orders is indicated by white color, buy/sell orders by red/blue, and the bid-ask spread by black. For every price level we indicated the total volume of limit orders by coloring as indicated on the right. The ends of the scale correspond to orders for $30000$ shares or more.}
\label{fig:pageshot2_year}
\end{figure*}

The time evolution of the order book during the whole year $2002$ is shown in Fig.
\ref{fig:pageshot2_year}. From this plot one gains little
additional information as compared to a plot of only the price. The
bid-ask spread, as noted before, is effectively negligible. One can also
clearly see that the shape of order book is -- on average -- symmetric between bids and asks.
The book contains a significant number of orders very far, 
sometimes $100$ ticks away from the best bid/ask. The width of the order book fluctuates in time, but one can see from Fig. \ref{fig:pageshot2_year} that it reverts to the typical value on the scale of weeks. This width can be calculated: on average, $50\%$ of the total limit order volume (both bid and ask) is within a $68$ tick range. This is comparable to the monthly absolute return, which is $1.4 \times 10^2$ ticks. 

\section{Daily time scale}
\label{sec:daily}

From previous analyses of TAQ data it is well known that daily returns look very different from monthly returns. In contrast with the ordinary diffusion picture of monthly data, daily returns are strongly fat tailed and day-to-day volatility autocorrelations are significant \cite{bouchaud.book}. On this scale there is also a negative correlation between past returns and future volatility: a price drop induces a greater excess volatility than an equal price increase \cite{bouchaud.leverage}. The bid-ask spread is still small, its mean value is $1.9$ ticks, while the daily absolute return is $23$ ticks on average. Note that, for trading strategies with the time horizon of a few days the spread is no longer negligible, because although it is small compared to volatility, it is no longer small as compared to their expected profits.

\begin{figure*}[ptb]
\centerline{\includegraphics[width=300pt,angle=-90]{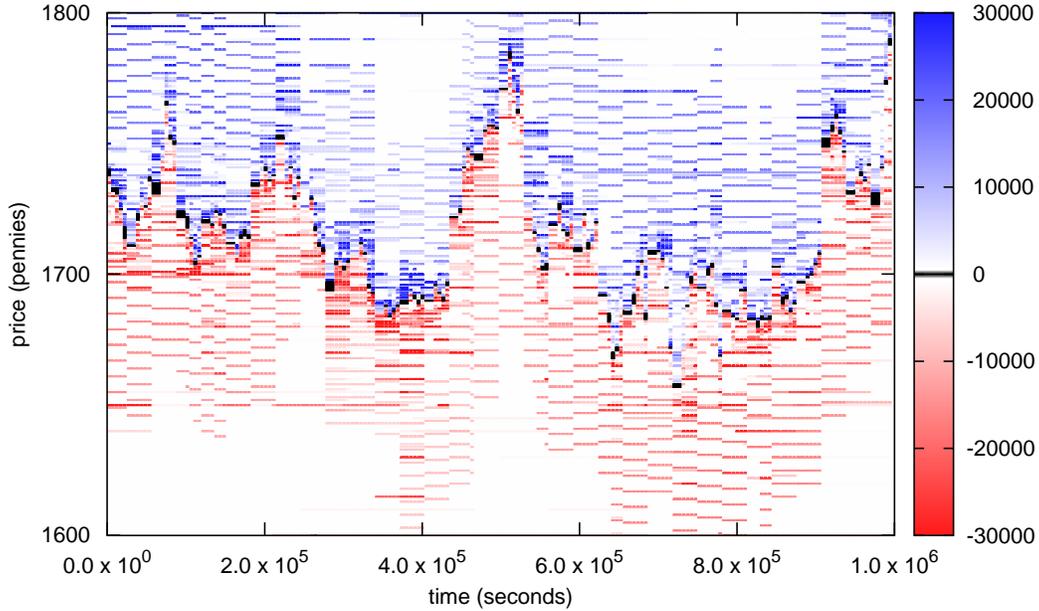}}
\caption{The order book of the stock GSK during the first $32$ trading days of $2002$. For all days during the period we made five snapshots of the order book at the $5000$'th, $10000$'th, $\dots$, $25000$'th second of the day. The lack of orders is indicated by white color, buy/sell orders by red/blue, and the bid-ask spread by black. For every price level we indicated the total volume of limit orders by coloring as indicated on the right. The ends of the scale correspond to orders for $30000$ shares or more.}
\label{fig:pageshot2_long3}
\end{figure*}

The trading day is a characteristic time in many respects. Trading activity and volatility
are known to follow intraday patterns, with typically higher values immediately
after the opening and before the closing of the market. Even more
importantly, for trading strategies a day is an important time horizon: day traders are required to close their positions at the end of the day, and even long-term strategies such as large investment portfolios are often adjusted daily. Consequently it is not surprising that most limit orders are canceled at the end of the trading day, and new limit orders are placed at the beginning of the next one.
The first $32$ trading days of the data period are shown in Fig. \ref{fig:pageshot2_long3}, where this daily cycle can be seen very clearly. The orders that remain in the book overnight are typically placed at technical levels \cite{osler.technical} constituted by round numbers such as $1700$ and $1650$ ticks in the first week in Fig. \ref{fig:pageshot2_long3}. 

\begin{figure*}[ptb]
\centerline{\includegraphics[width=200pt,angle=-90,trim=110 0 0 0]{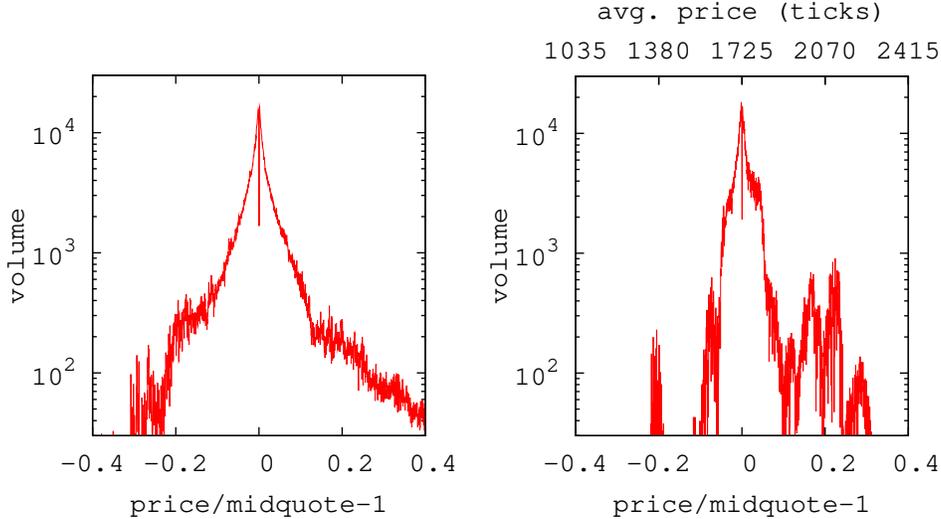}}
\vskip5mm
\caption{\emph{Left}: The average shape of the order book of GSK calculated during the year $2002$. The prices were then divided by the current midquote price, and we subtracted $1$. \emph{Right}: The average shape of the order book of GSK calculated during the first $10$ trading days of the year $2002$. The indicated average prices correspond to $1725 \times (\mathrm{price/midquote}-1)$ ticks. \emph{Note}: Averages were calculated by taking snapshots of the book during every trading day of the period from its $3000$'th to its $28000$'th second every $1000$ seconds. The prices were then divided by the current midquote price, and we subtracted $1$.}
\label{fig:shapeavg}
\end{figure*}

\begin{figure*}[ptb]
\centerline{\includegraphics[width=95pt,angle=-90,trim=200 60 60 130]{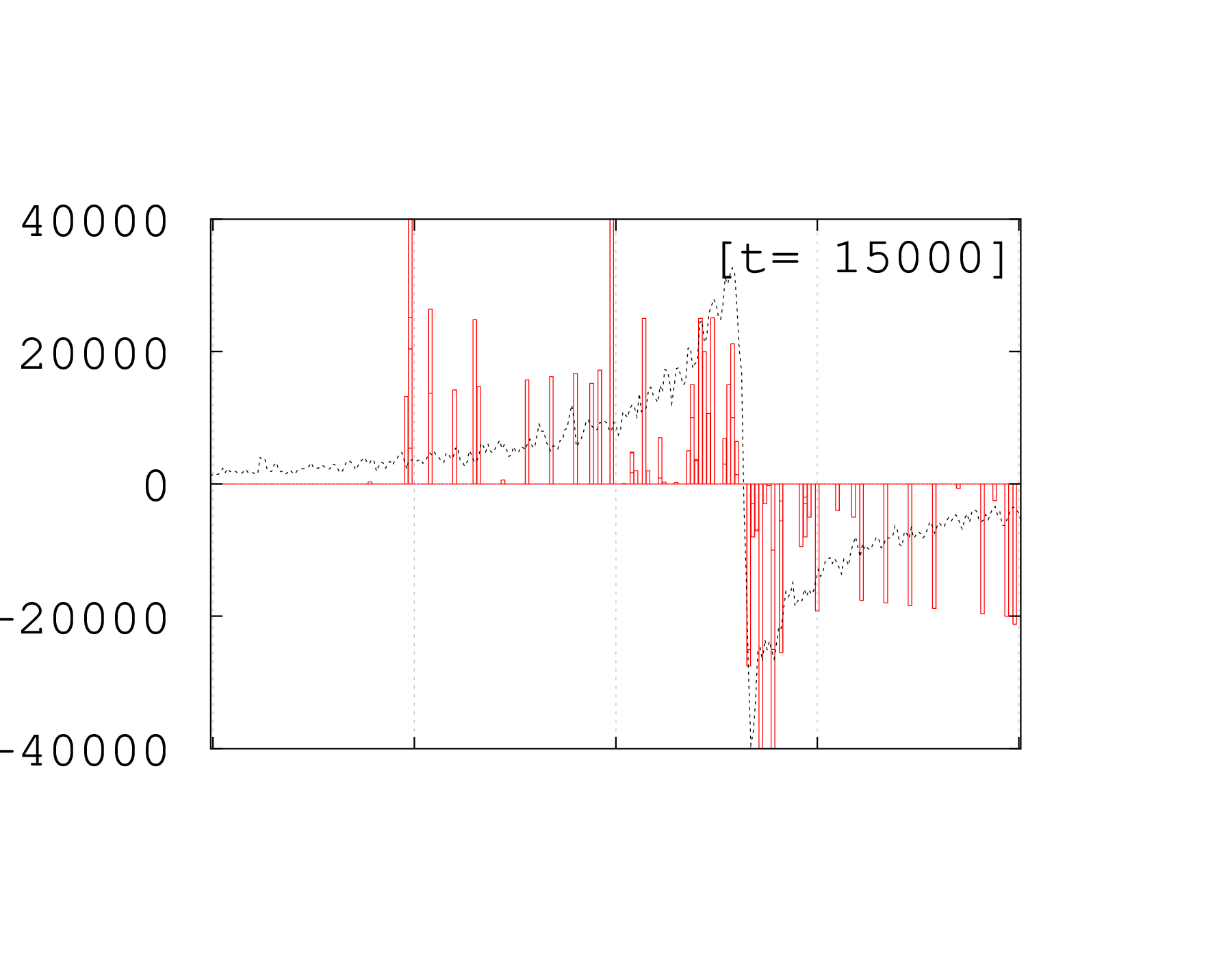}\includegraphics[width=95pt,angle=-90,trim=200 60 60 130]{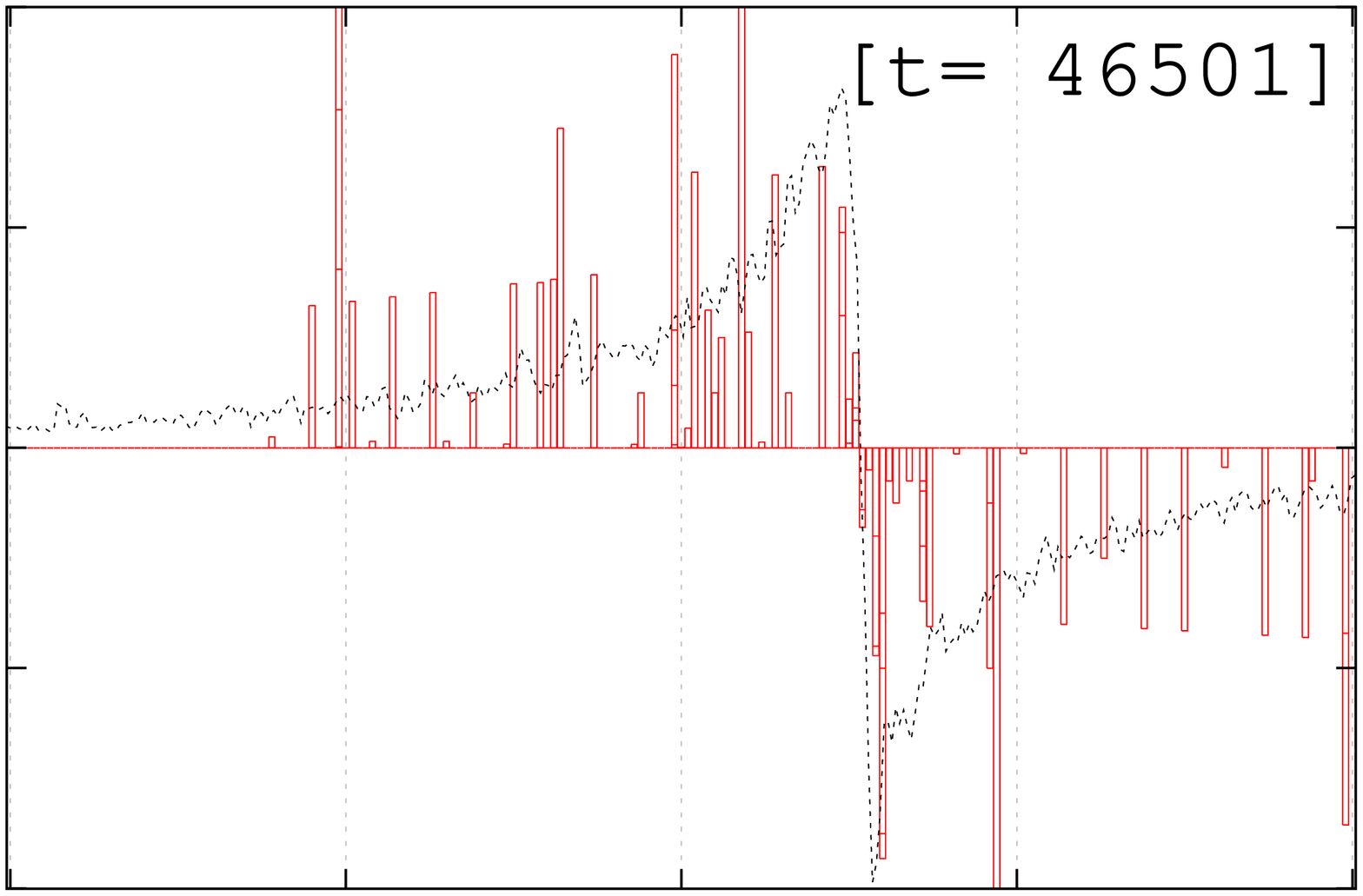}\includegraphics[width=95pt,angle=-90,trim=200 60 60 130]{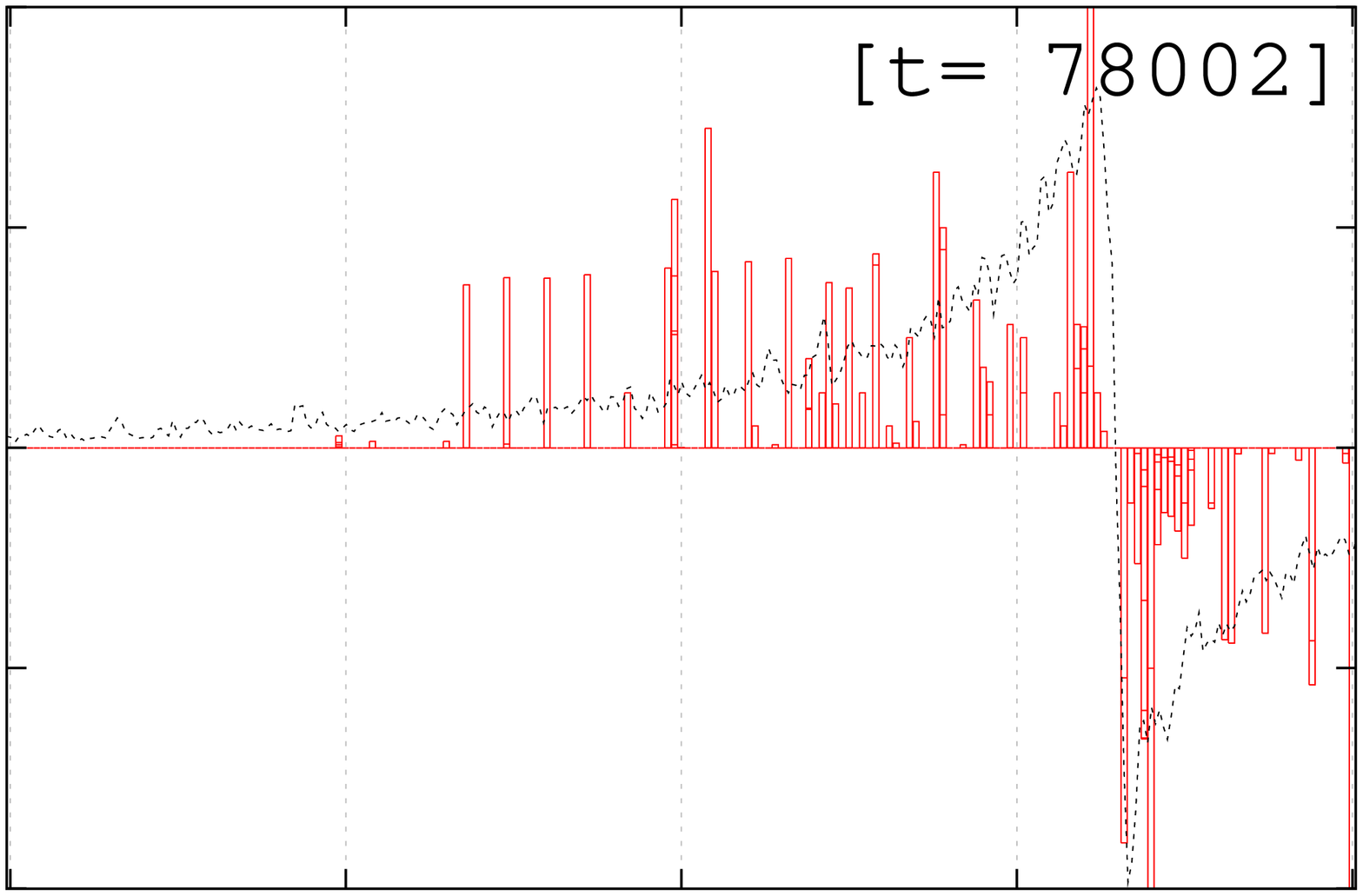}}
\centerline{\includegraphics[width=95pt,angle=-90,trim=200 60 60 130]{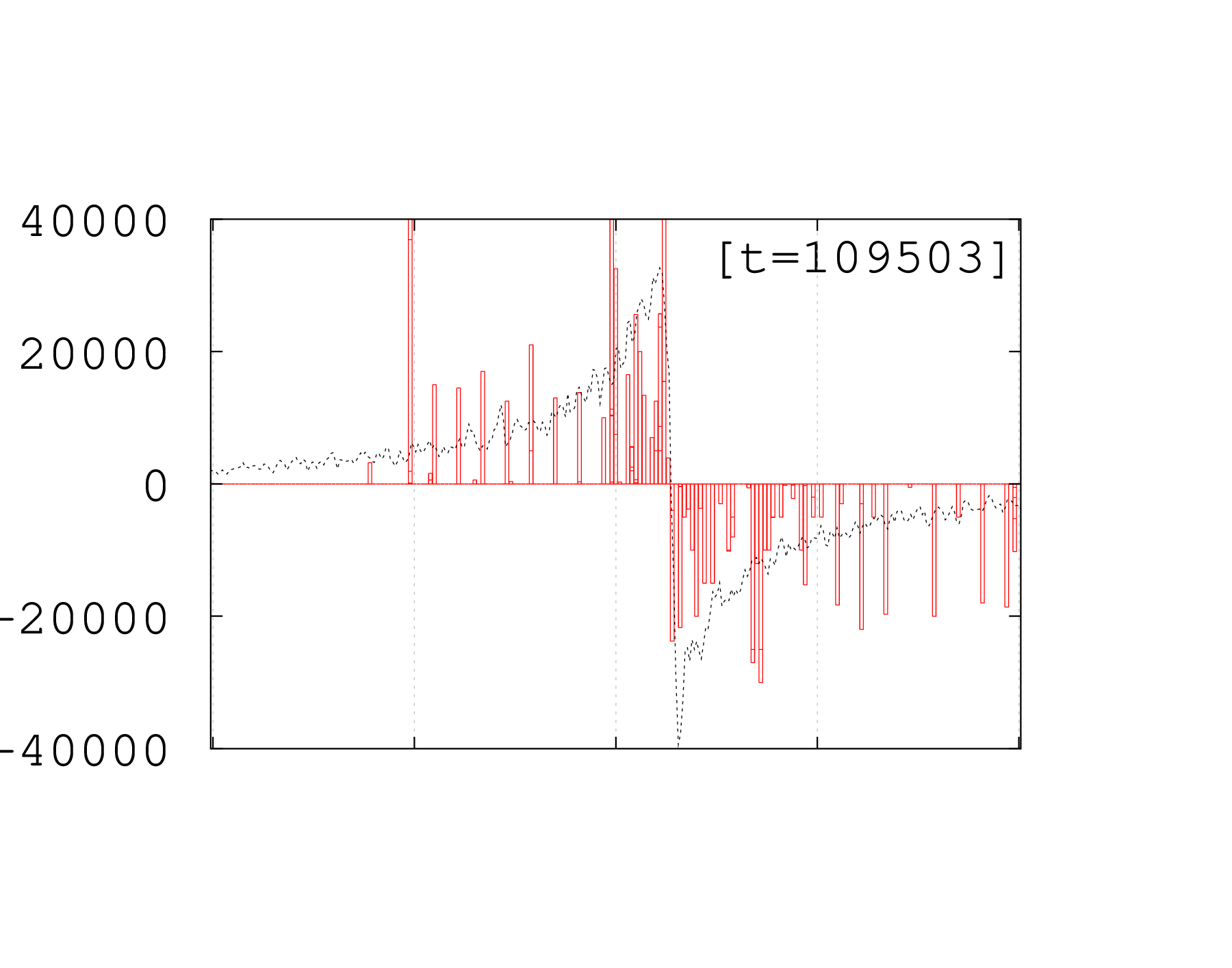}\includegraphics[width=95pt,angle=-90,trim=200 60 60 130]{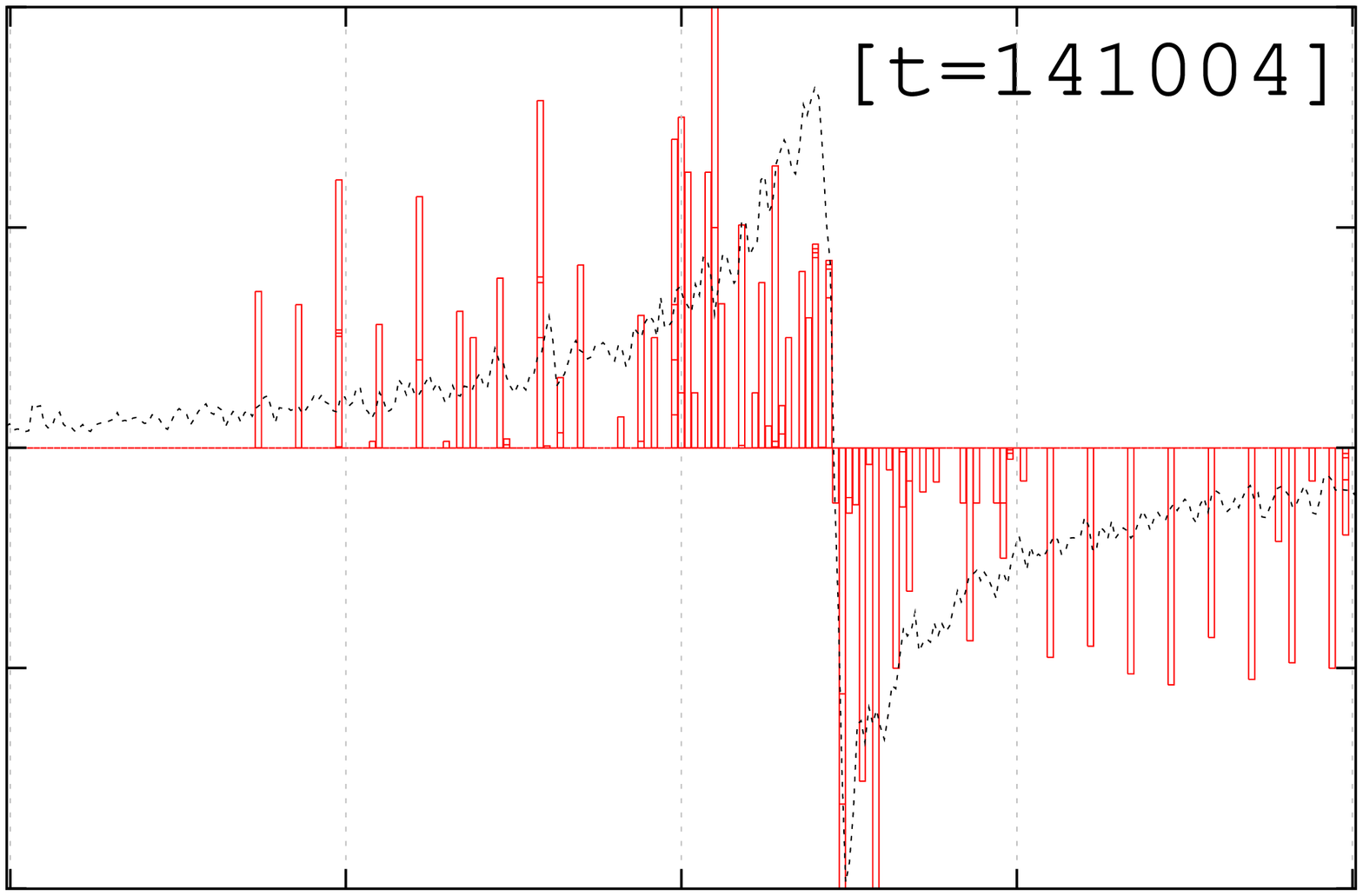}\includegraphics[width=95pt,angle=-90,trim=200 60 60 130]{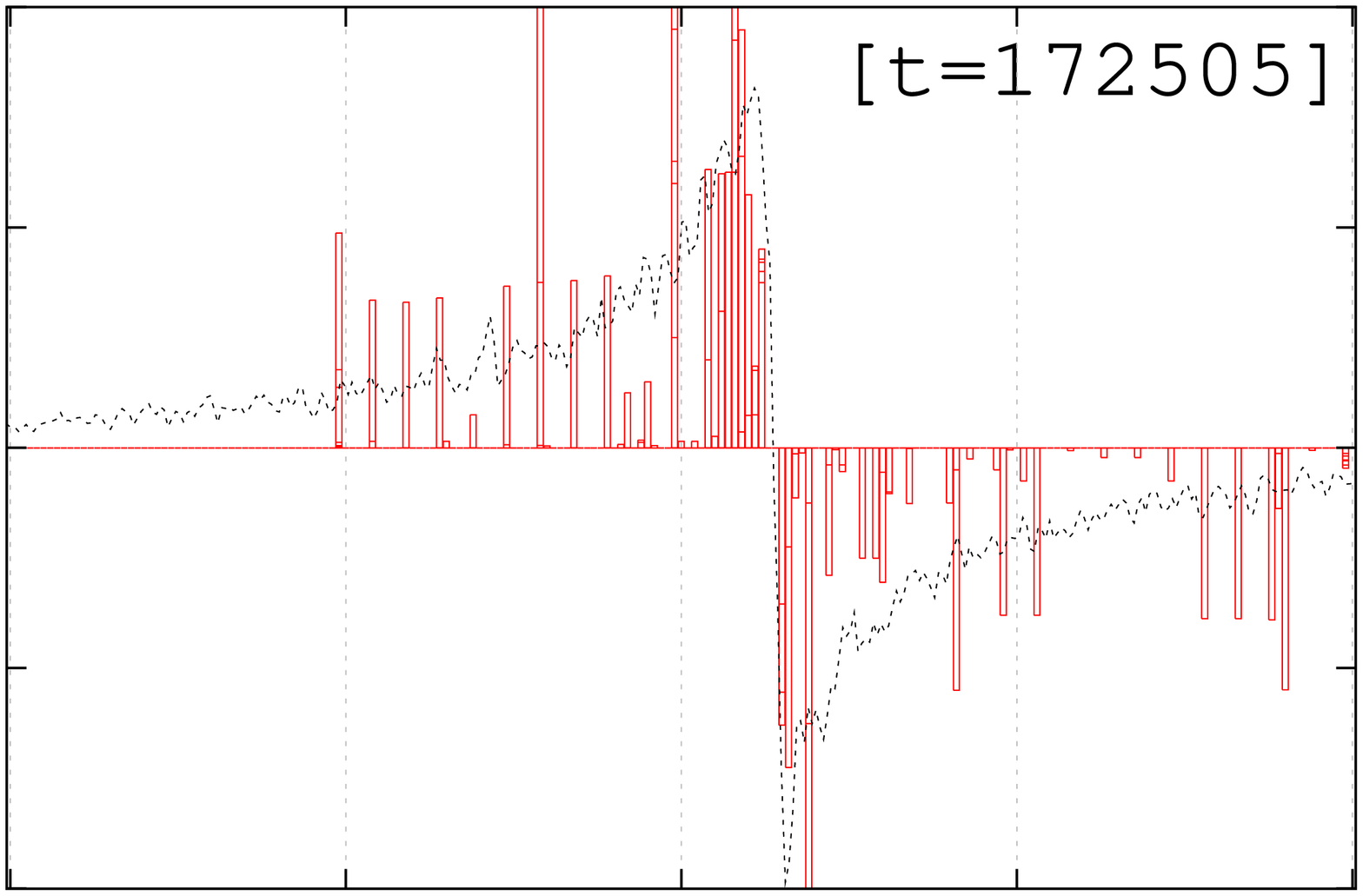}}
\centerline{\includegraphics[width=95pt,angle=-90,trim=200 60 60 130]{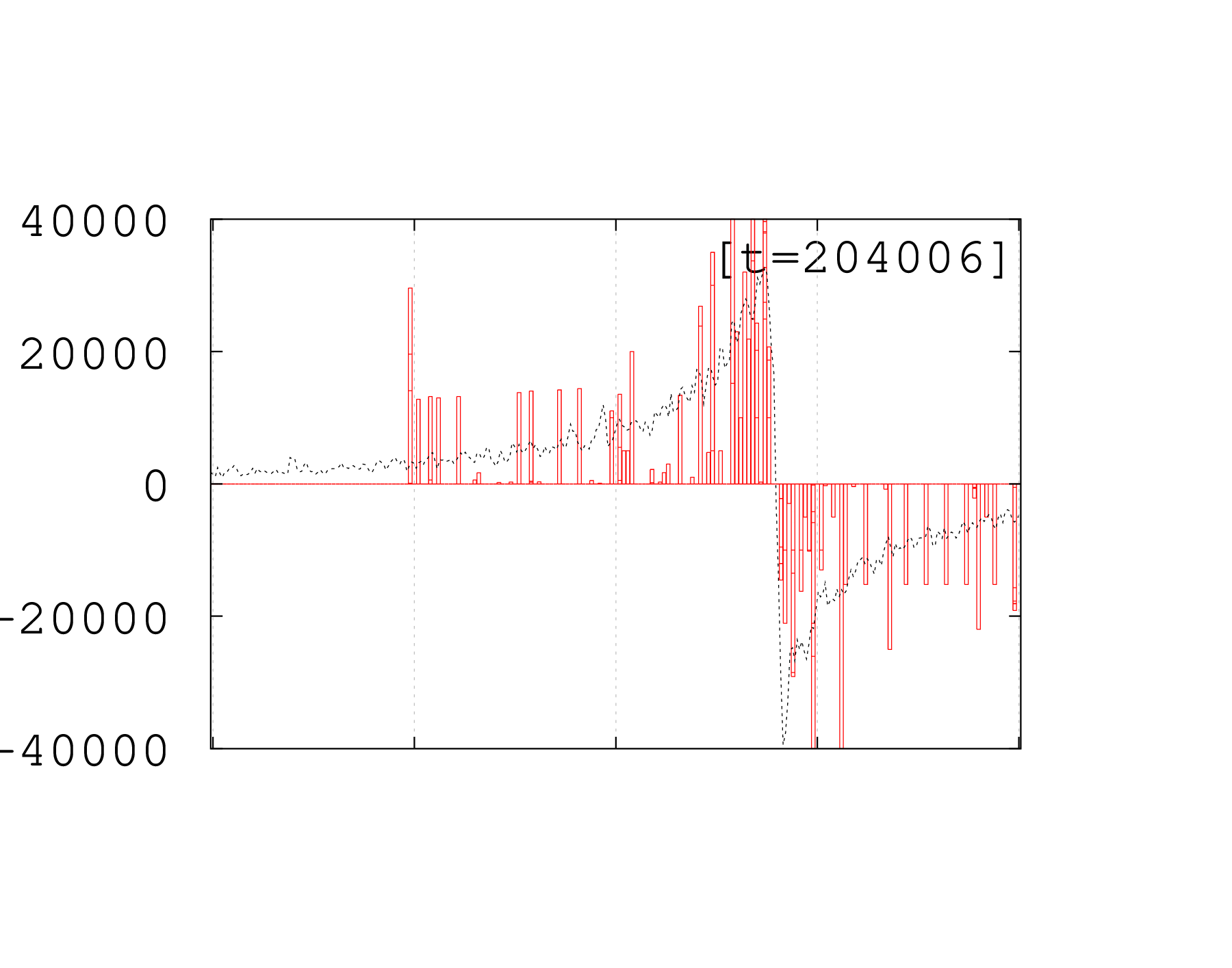}\includegraphics[width=95pt,angle=-90,trim=200 60 60 130]{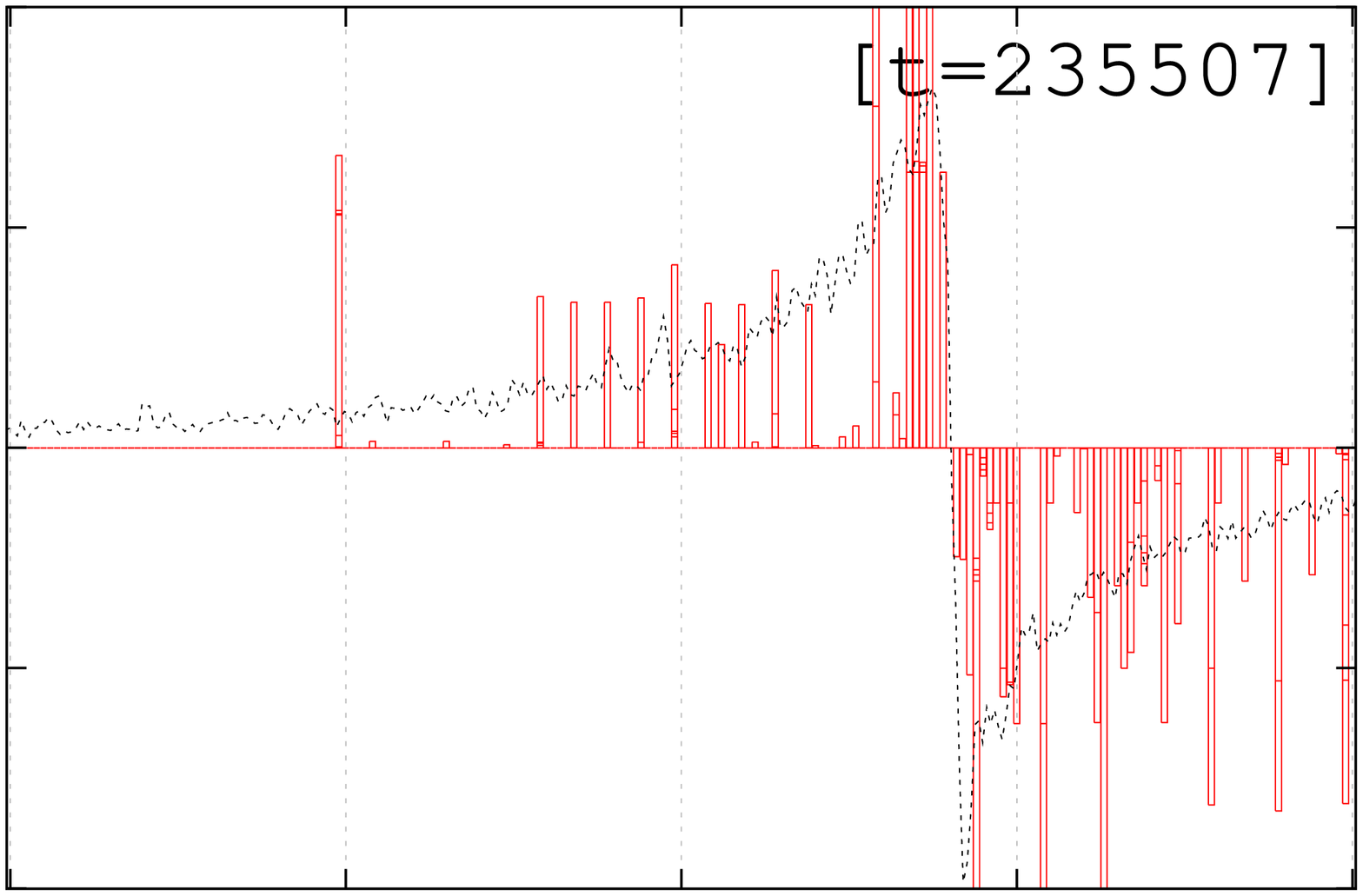}\includegraphics[width=95pt,angle=-90,trim=200 60 60 130]{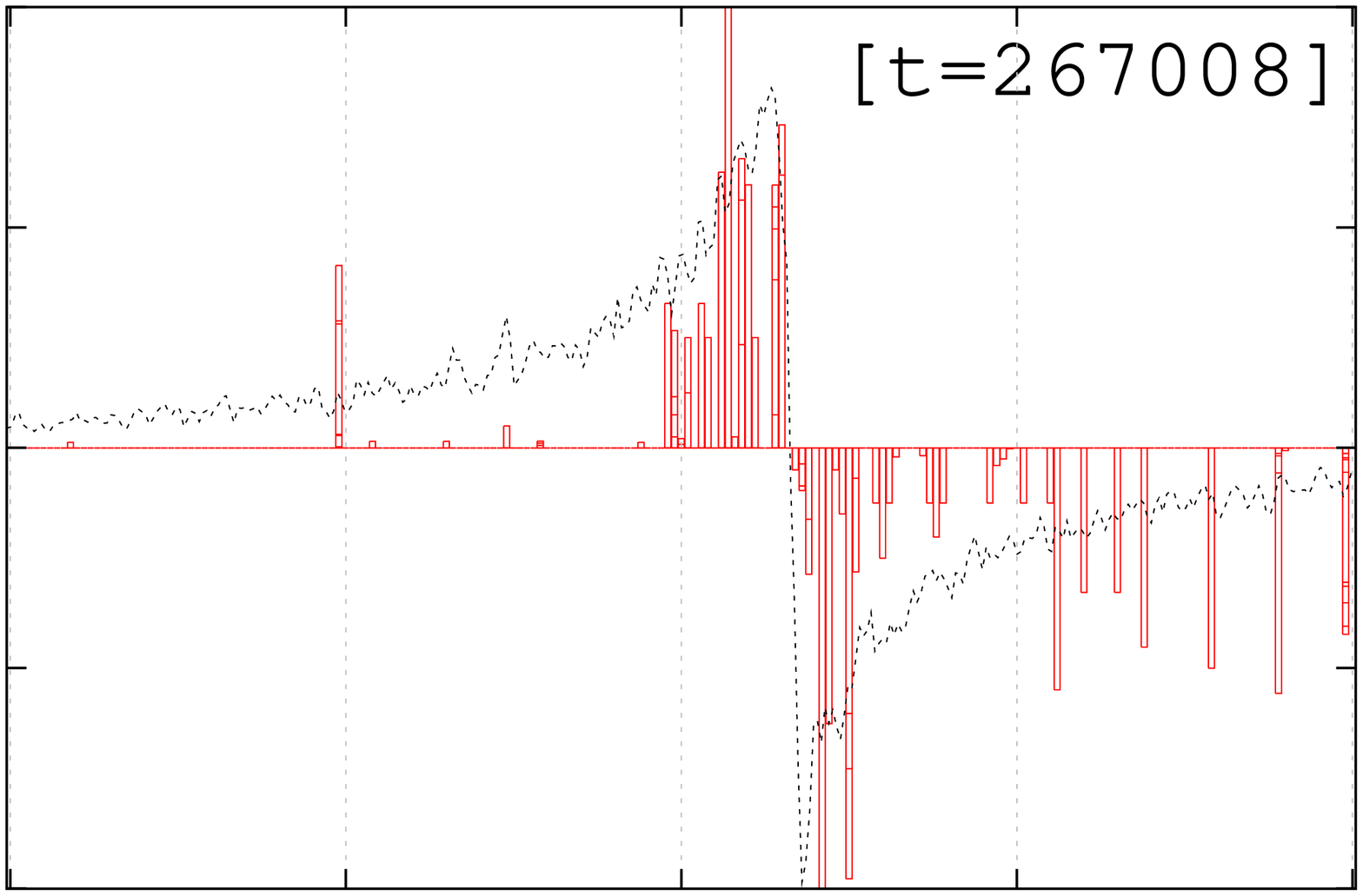}}
\centerline{\includegraphics[width=95pt,angle=-90,trim=200 60 60 130]{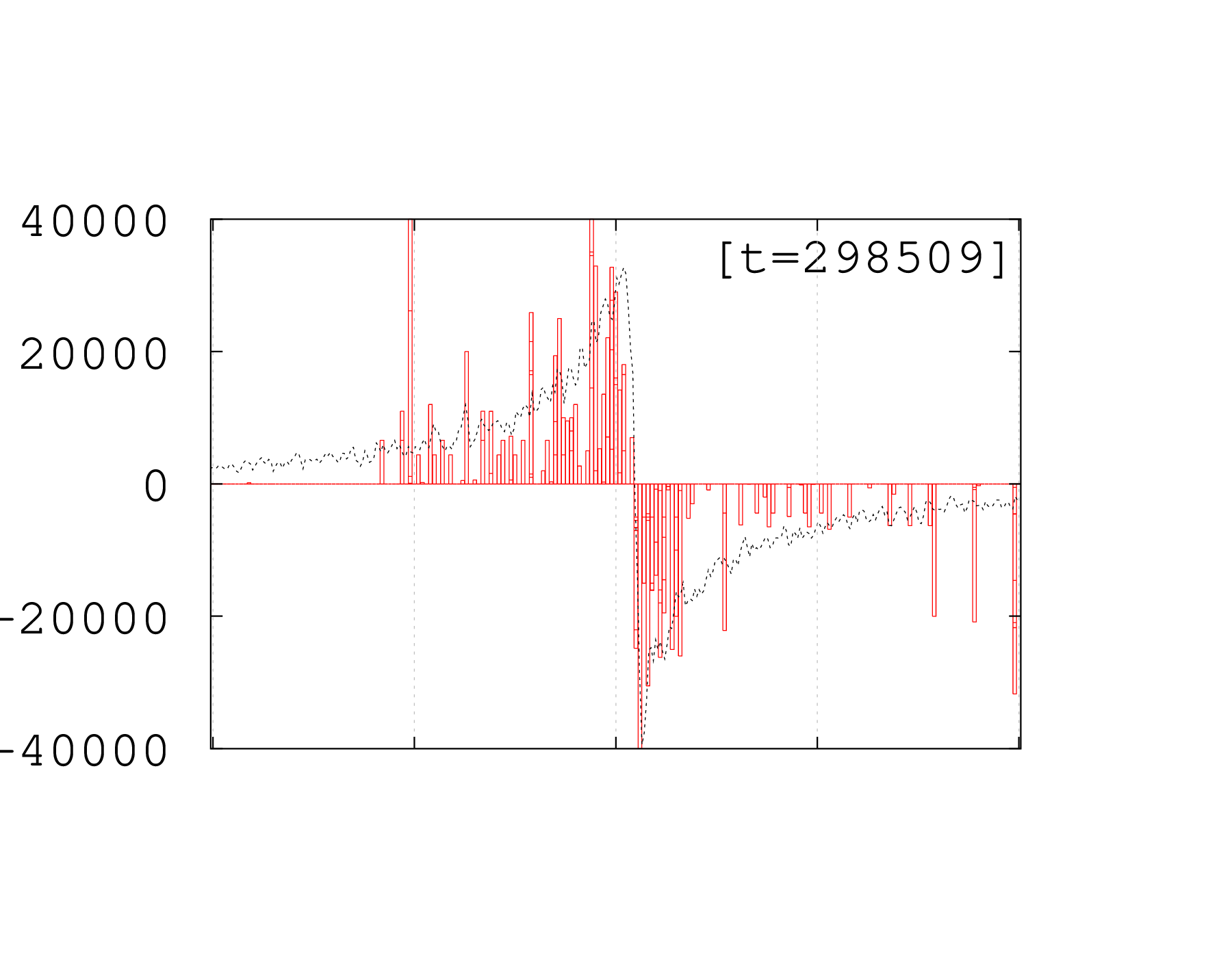}\includegraphics[width=95pt,angle=-90,trim=200 60 60 130]{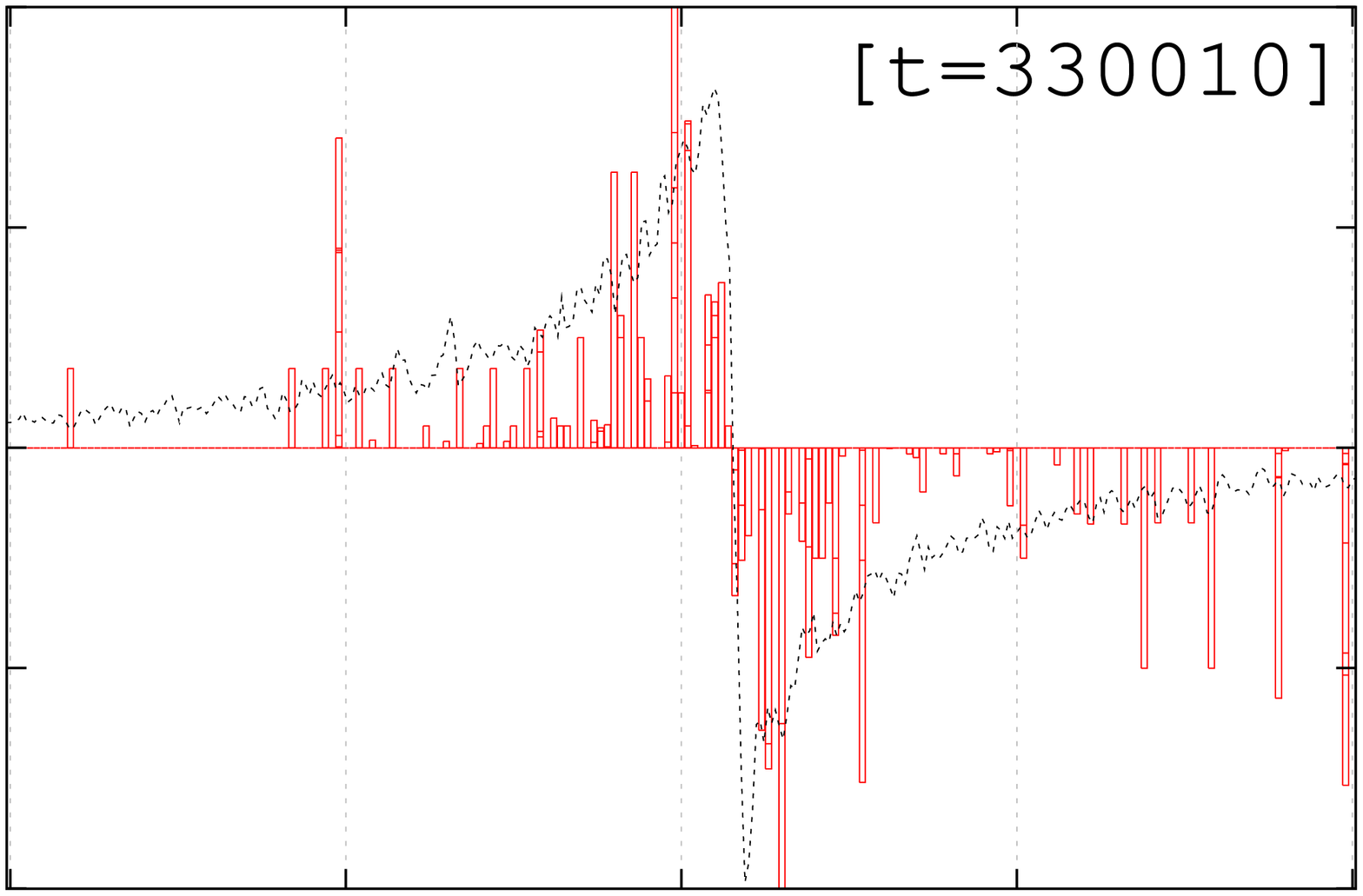}\includegraphics[width=95pt,angle=-90,trim=200 60 60 130]{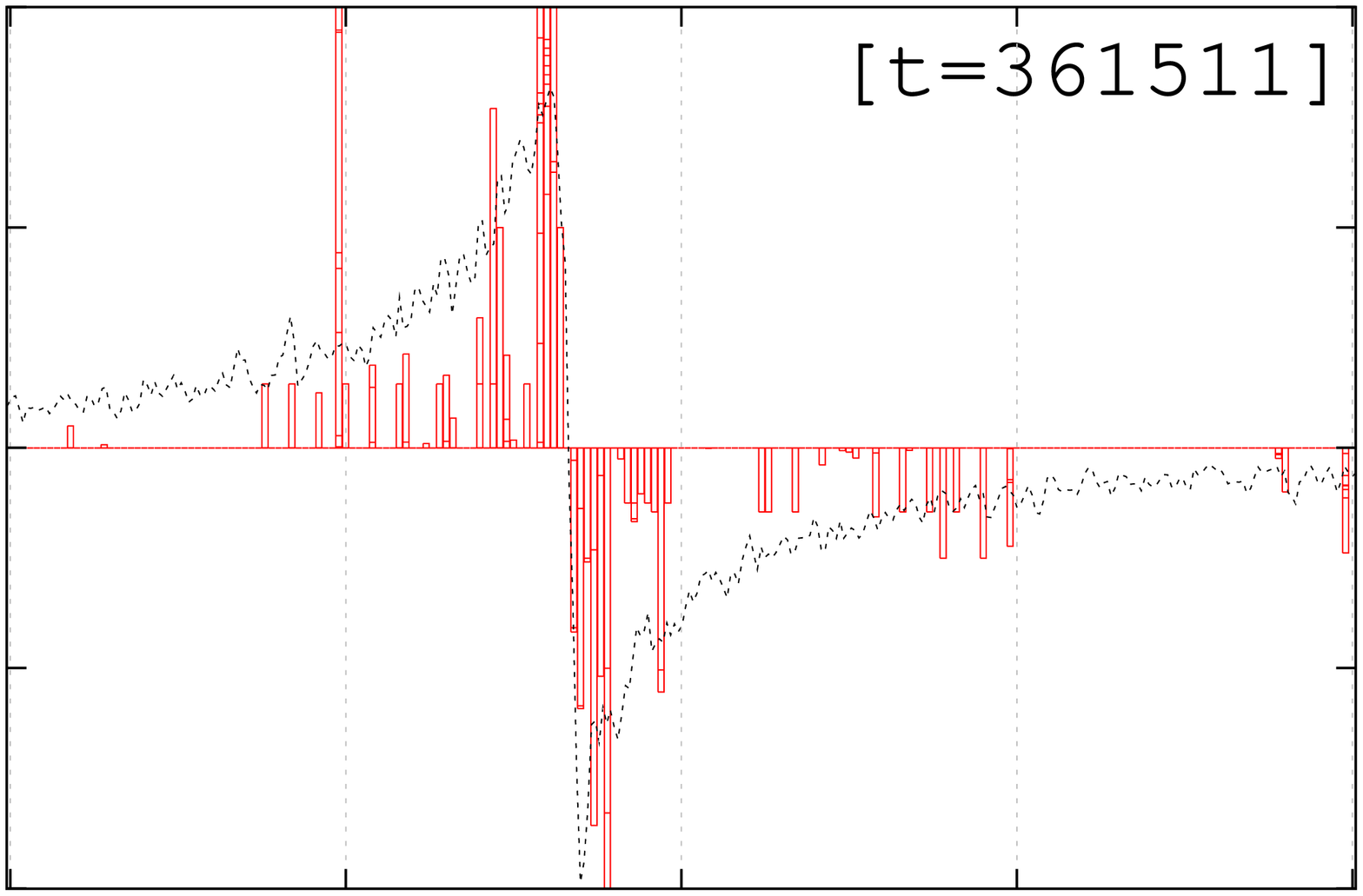}}
\centerline{\includegraphics[width=95pt,angle=-90,trim=200 60 60 130]{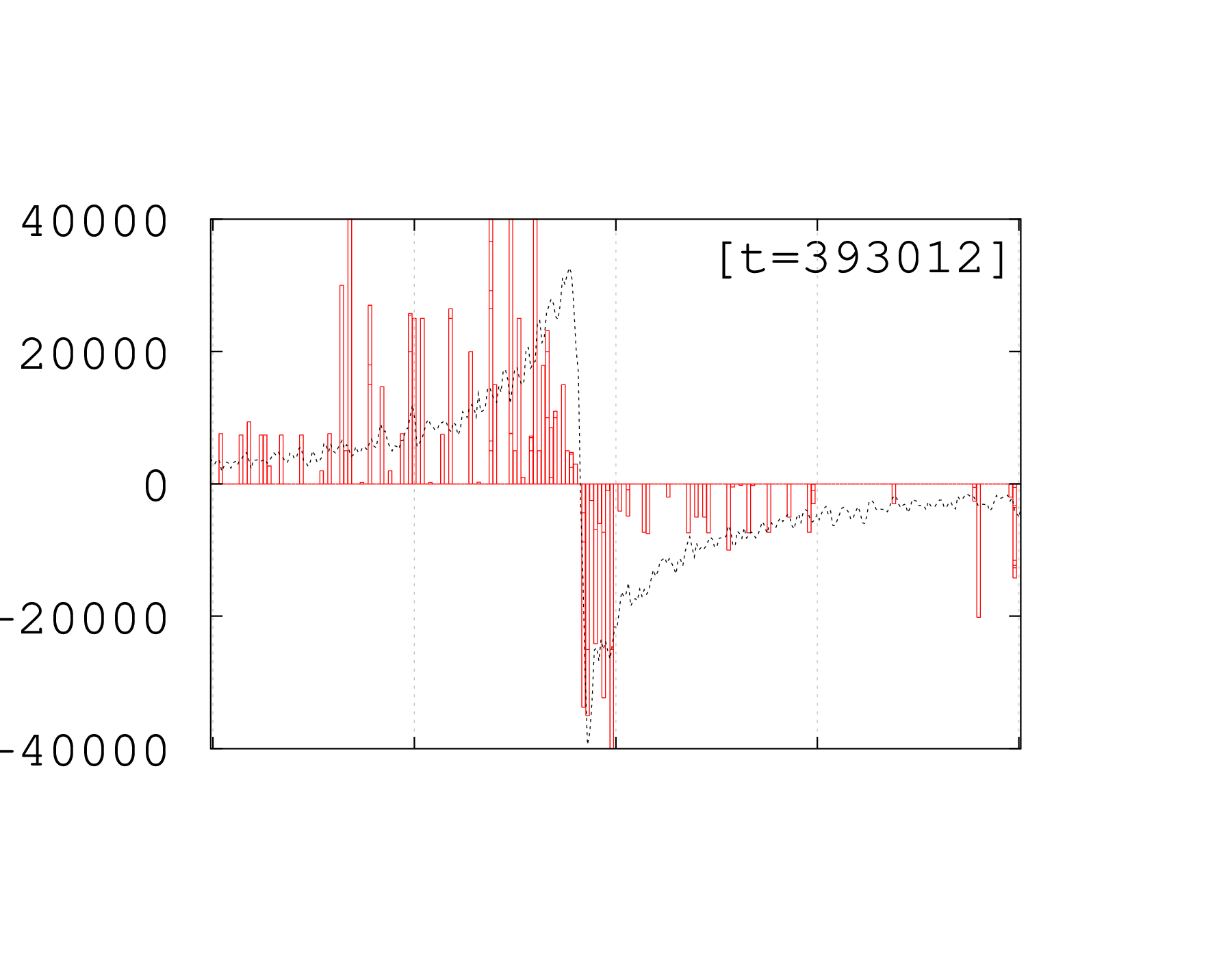}\includegraphics[width=95pt,angle=-90,trim=200 60 60 130]{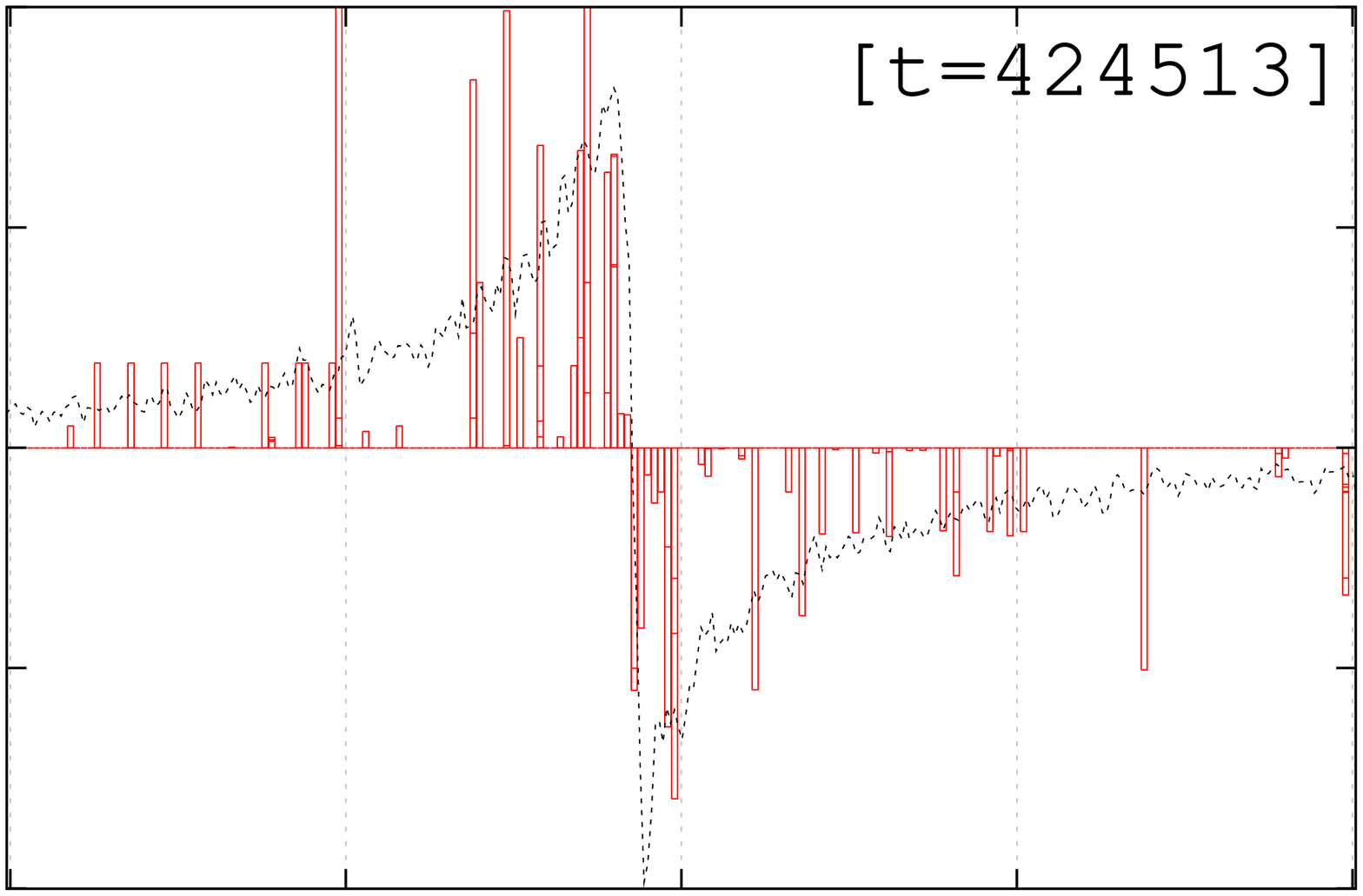}\includegraphics[width=95pt,angle=-90,trim=200 60 60 130]{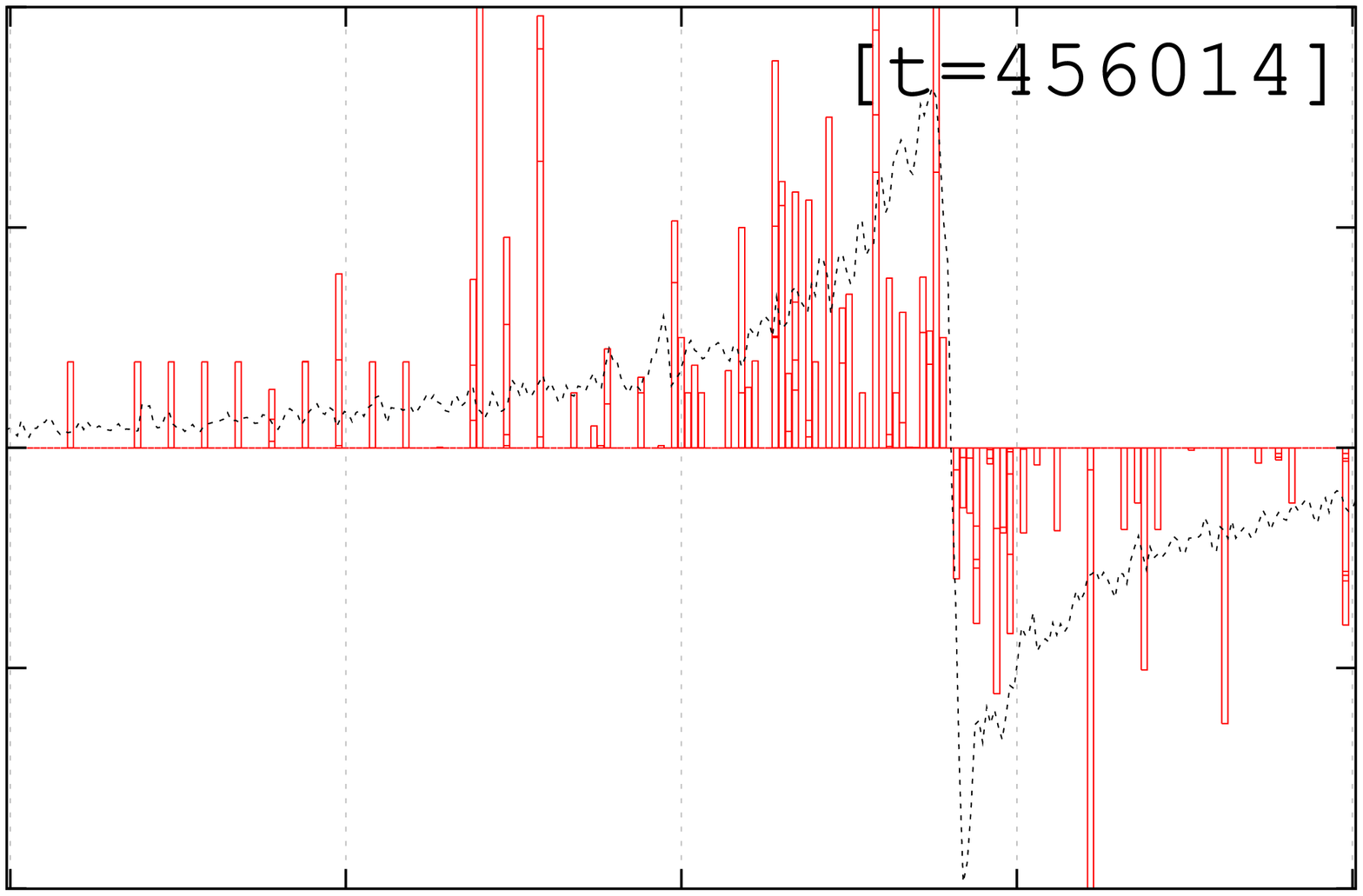}}
\centerline{\includegraphics[width=95pt,angle=-90,trim=200 60 60 130]{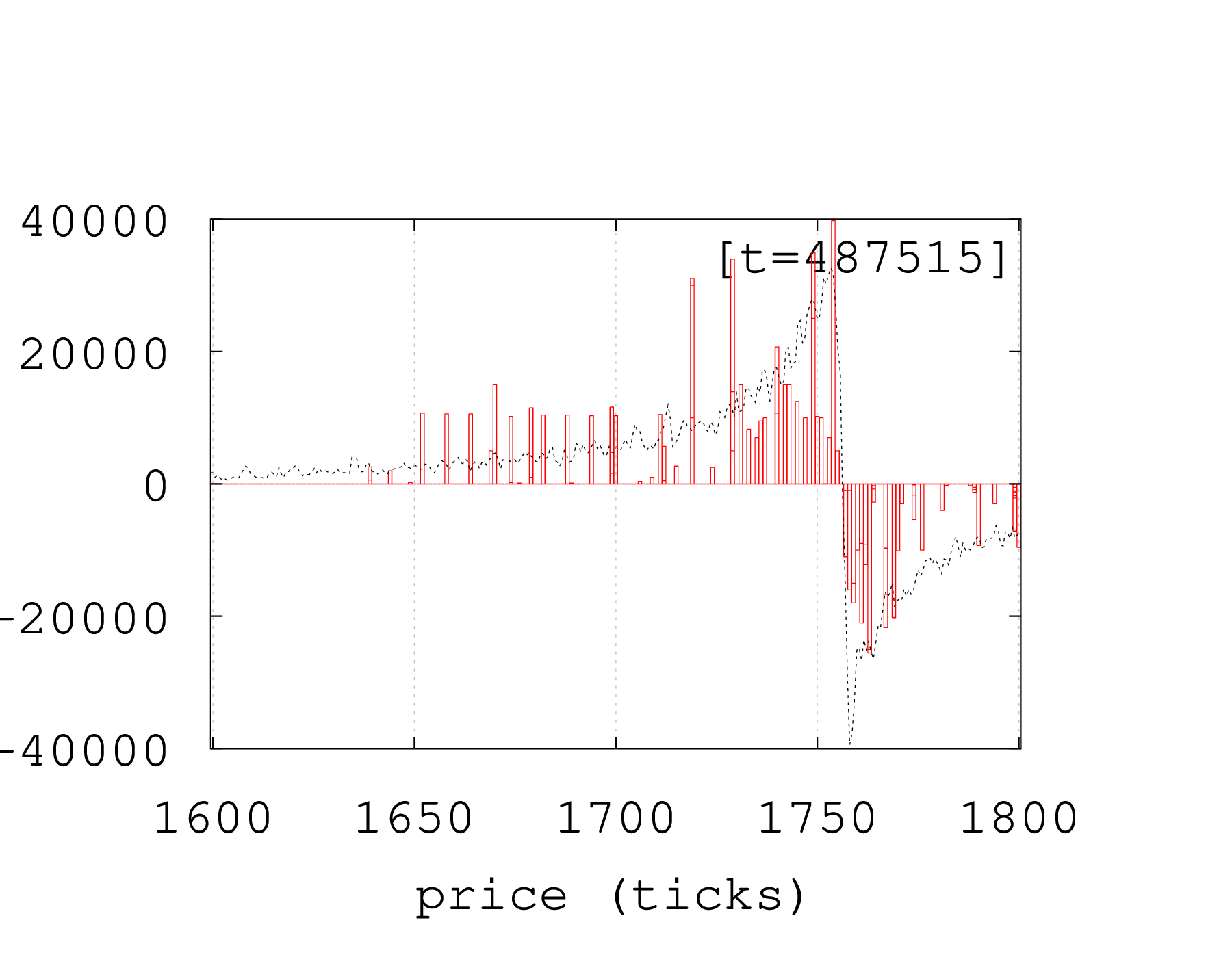}\includegraphics[width=95pt,angle=-90,trim=200 60 60 130]{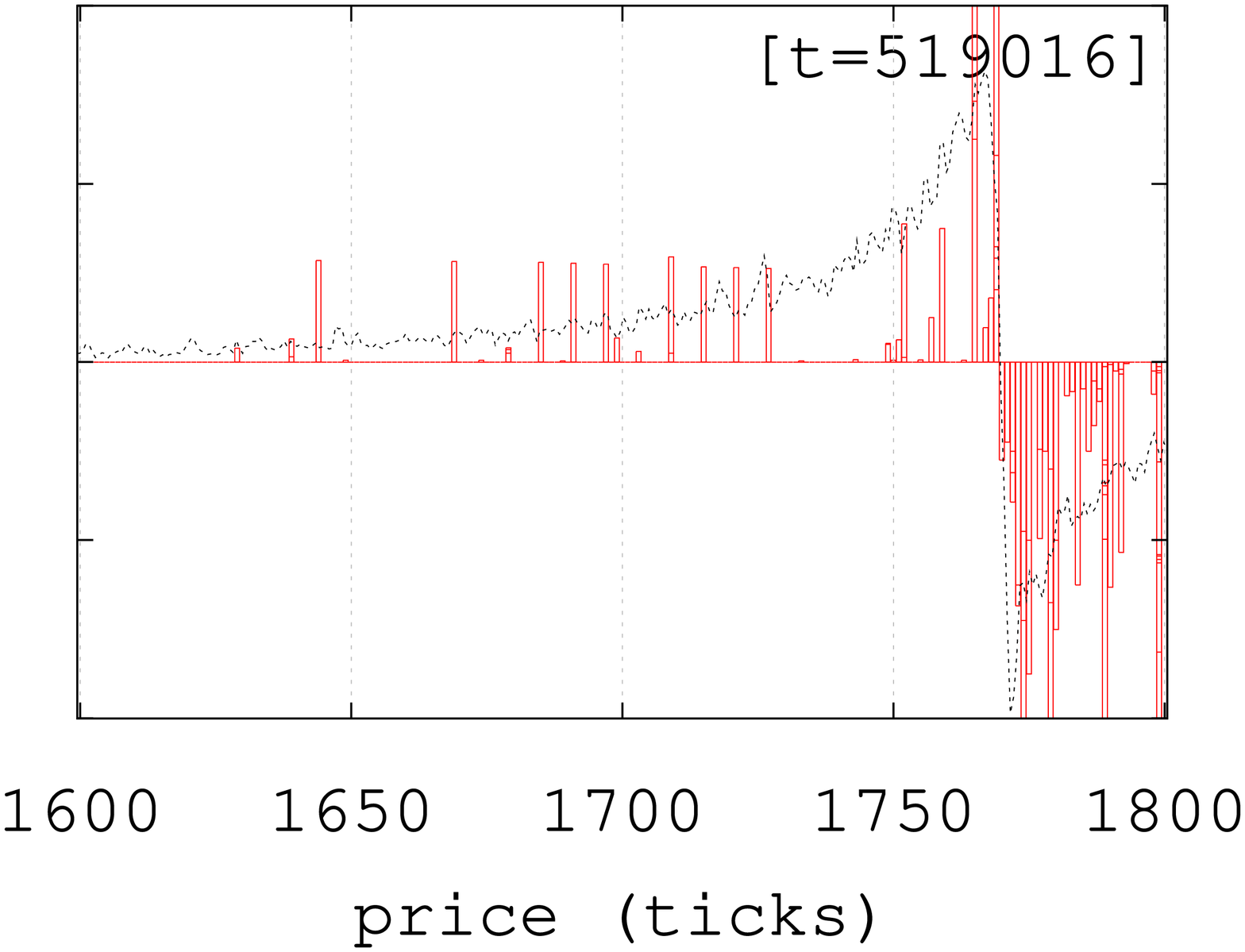}\includegraphics[width=95pt,angle=-90,trim=200 60 60 130]{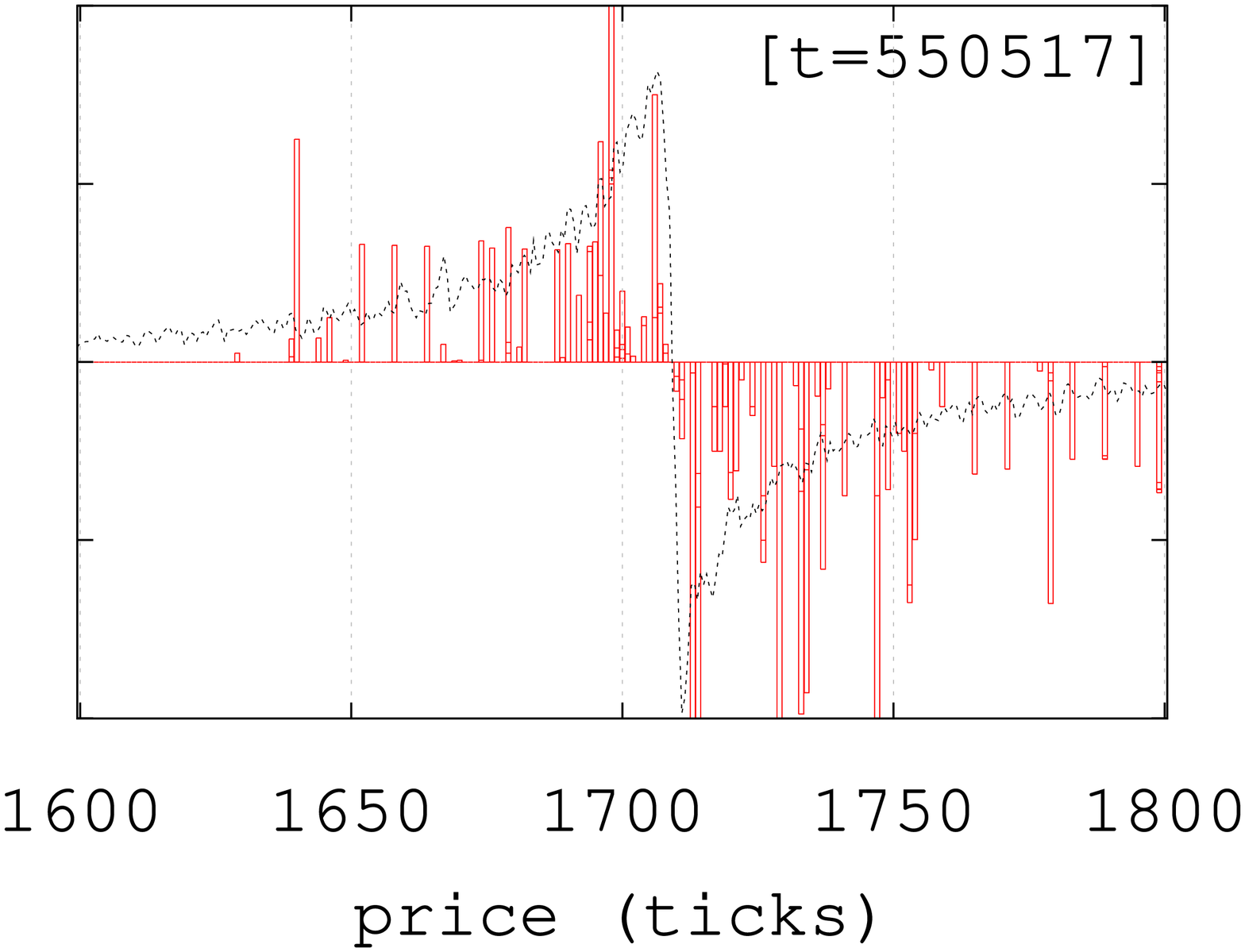}}
\vskip5mm
\caption{The order book of the stock GSK during the first $18$ trading days of $2002$. For all days we made snapshots of the order book at the $15000$'th second of the day. The length of the bars indicates the total volume of the orders at that price level, the maximum length corresponds to volumes for $40000$ shares or more. The dotted line corresponds to the time average of the book shape at the $15000$'th second of trading days. The shape is similar, but the amplitude is somewhat larger as in Fig. \ref{fig:shapeavg}(left).}
\label{fig:pm3d_long3}
\end{figure*}

Previous studies have reported a slow, continuous, symmetric decay of
limit order volume as a function of the distance from the spread (see, e.g., Ref. \cite{potters.pa2003}). While this true on average, it is very uncharacteristic of the state of the
book at any given time. This difference has predominantly the following two origins:

(i) In Fig. \ref{fig:shapeavg}(left) we plot the average shape of the order book over the whole year. Then in Fig. \ref{fig:pm3d_long3} the shape in the $15000$'th second of the first $18$ trading days is compared with the average pattern\footnote{In fact the average shape of the book shows some variation during the trading day. Because Fig. \ref{fig:pm3d_long3} displays snapshots taken at the $15000$'th second of trading days, the average was also calculated over such snapshots.}. The shape of the order book varies strongly from day to day, but it is always very different from the average. In other words, the \emph{average} shape is not \emph{typical}. For example, Fig. \ref{fig:shapeavg}(right) shows the shape of the book as an average over the first $10$ days of our data. Even such a time average looks very different from the yearly average.

(ii) From Fig. \ref{fig:pageshot2_long3} one can see that the book consists of three types of orders. 
(a) There is a narrow range around the bid-ask spread where there is a high concentration of limit orders carrying a substantial fraction of the volume in the book. (b) Fig. \ref{fig:pageshot2_long3} is also very "striped" along the price axis: there is a regular pattern of moderately sized orders every $5$ ticks to a distance of up to $100$ ticks from the spread. The same can be seen in most snapshots in Fig. \ref{fig:pm3d_long3}. (c) In addition there appear to be much larger limit orders placed around round numbers, such as $1700$ and $1650$ ticks in this example (see Fig. \ref{fig:pageshot2_year}).

The central part (a) will be addressed in the next section. The regularly striped pattern (b) contains relatively smaller volumes, and it does not show up at all in the yearly or even the $10$-day average book shape in Fig. \ref{fig:shapeavg}. The reason is that most of the orders constituting 
pattern (b) remain unchanged during the whole trading day, and hence they continuously change position
relatively to the midquote price. Thus a time average in which distance is calculated from the midquote will only indicate a continuous pattern. Also notice that the pattern is not fixed to round numbers (quotes ending to $5$ or $0$, if the price moves, the pattern is replaced at different levels on the next trading day.

Finally the orders at the round levels (c) are relatively large. They do not appear in the yearly average book shape, but some are large and stable enough to be seen in, e.g., the $10$-day average in Fig. \ref{fig:shapeavg}(right). The average was made in a period when the midquote price fluctuated around $1725$ ticks, and so an approximate price axis could also be added to the plot. There are several clear peaks corresponding to limit orders that are very far from the spread.

\begin{figure*}[ptb]
\centerline{\includegraphics[width=320pt,angle=-90]{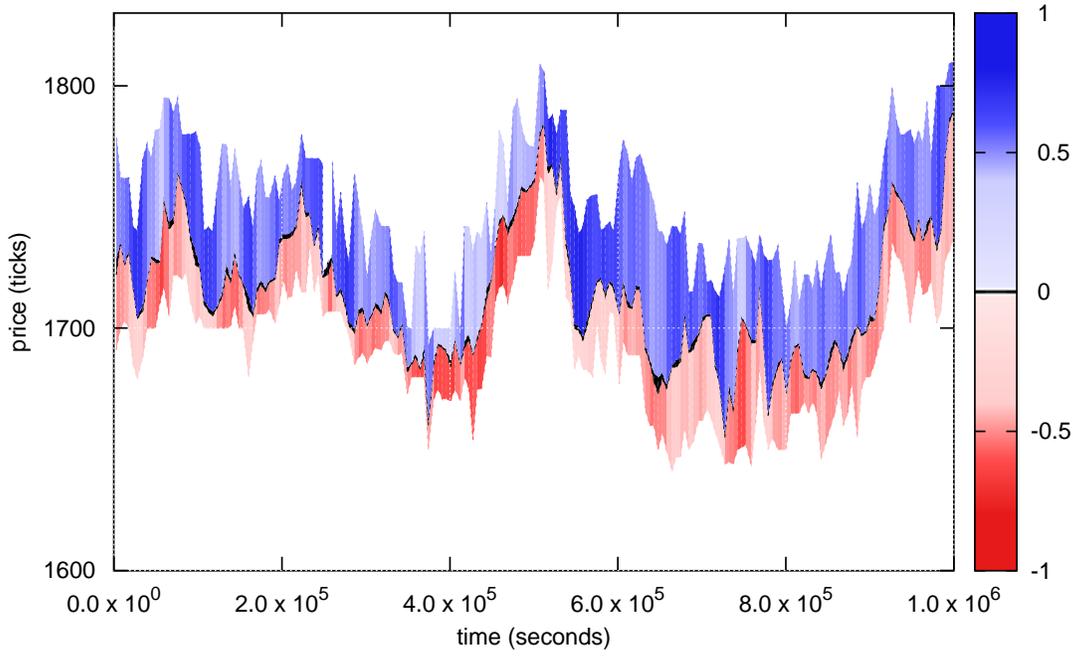}}
\caption{The order imbalance and the median-to-median width of the order book of GSK calculated during the first $10$ trading days of the year $2002$. We took six snapshots of the book during every trading day of the period from its $3000$'th to its $28000$'th second every $5000$ seconds. Red/blue color corresponds to buy/sell orders, and the black line between them to the bid-ask spread.	The coloring also indicates the value of $I_t^\mathrm{buy}$ and $I_t^\mathrm{sell}$ (see right for values), while the width of the graph shows the median of the book in terms of its total volume.}
\label{fig:pageshape_long3}
\end{figure*}

From Fig. \ref{fig:pm3d_long3} one can see that not only the total volume inside the book fluctuates, but there is always an asymmetry present between the buy and sell sides, whose degree and direction varies in time. This imbalance can be quantified, for example, as follows. Let us denote the buy/sell volume
at price $p$ by $v^\mathrm{buy/sell}_t(p)$. The total buy/sell volume 
is given by
$$V_t^\mathrm{buy}=\sum_{p=0}^{b_t}v_t^\mathrm{buy}(p)$$
and
$$V_t^\mathrm{sell}=\sum_{p=a_t}^{\infty}v_t^\mathrm{sell}(p).$$
We define the buy/sell imbalance as
$$I_t^\mathrm{buy}=\frac{V_t^\mathrm{buy}}{V_t^\mathrm{buy}+V_t^\mathrm{sell}}$$
and
$$I_t^\mathrm{sell}=-\frac{V_t^\mathrm{sell}}{V_t^\mathrm{buy}+V_t^\mathrm{sell}}.$$
Trivially $I_t^\mathrm{buy}-I_t^\mathrm{sell}\equiv 1$. The behavior of the imbalance is plotted in Fig. \ref{fig:pageshape_long3}. The relationship between daily return and imbalance is less than straightforward. For example, there is a several day period around $t = 4-5 \times 10^5$, where there is a consistent upward price trend combined with an long lasting excess of buy limit orders. The pattern is then repeated downwards for $t=5-7\times 10^5$. On the other hand, in the period $t=8-9\times 10^5$ there is an upward trend and still there are excess sell limit orders. The lack of direct causality between limit order volume and returns is of course not very surprising. The efficient nature of the market would not allow such a simple predictor of returns to exist, but the relationship between imbalance, returns and predictability is a question still under debate \cite{farmer.efficiency, weber.qf2005, bouchaud.molasses, farmer.whatreally}.

\section{Intraday and tick-by-tick time scales}
\label{sec:intraday}

Finally, let us look at intraday and tick-by-tick data. Again, there is a lot of recent knowledge about the statistics of short time returns. They are known to be broadly distributed, the tails of the distribution often being fitted with a power law decay \cite{lux.paretian, gopi.inversecube}. Volatility tends to be highly correlated in intraday data, but the signs of returns are essentially unpredictable beyond a few minutes \cite{bouchaud.book}. The average time between consecutive trades was $10$ sec for GSK in particular, but it is always of the order of seconds for any liquid stock or market \cite{eisler.orderbook, ivanov.itt}.

\begin{figure*}[ptb]
\centerline{\includegraphics[width=300pt,angle=-90]{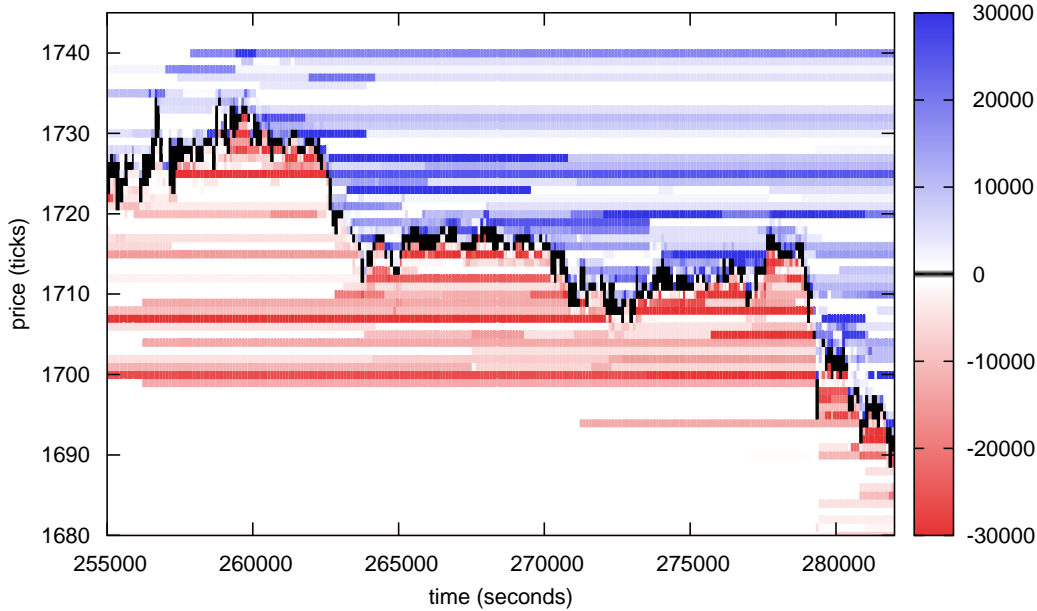}}
\caption{The order book of the stock GSK during the $8$'th trading day of $2002$. During the period a snapshots of the order book every $50$ seconds. The lack of orders is indicated by white color, buy/sell orders by red/blue, and the bid-ask spread by black. For every price level we indicated the total volume of limit orders by coloring as indicated on the right. The ends of the scale correspond to orders for $30000$ shares or more.}
\label{fig:pageshot2_short8}
\end{figure*}

The nature of the price formation process only becomes crucial on the very short run. The $8$'th day of trading is shown in Fig. \ref{fig:pageshot2_short8} from the order book perspective. For time horizons longer than $5-15$ minutes any liquid stock is traded tens-hundreds of times, and there is evidence, that the qualitative picture of anomalous diffusion for prices describes limit order executions well \cite{eisler.orderbook}. The typical time scale when the mean absolute return exceeds the mean bid-ask spread is also similar: The mean $1$ and $10$-minute absolute returns are $0.9$ and $3$ ticks, respectively, while the mean bid-ask spread is $1.9$ ticks.

As seen in Fig. \ref{fig:pm3d_long3}, during usual trading there is a regime of up to $10$ occupied levels immediately near the best prices. The regular striped pattern every $5$ ticks can is clearly present and during the day the levels are constant. 

\begin{figure*}[ptb]
\centerline{\includegraphics[width=260pt,angle=-90]{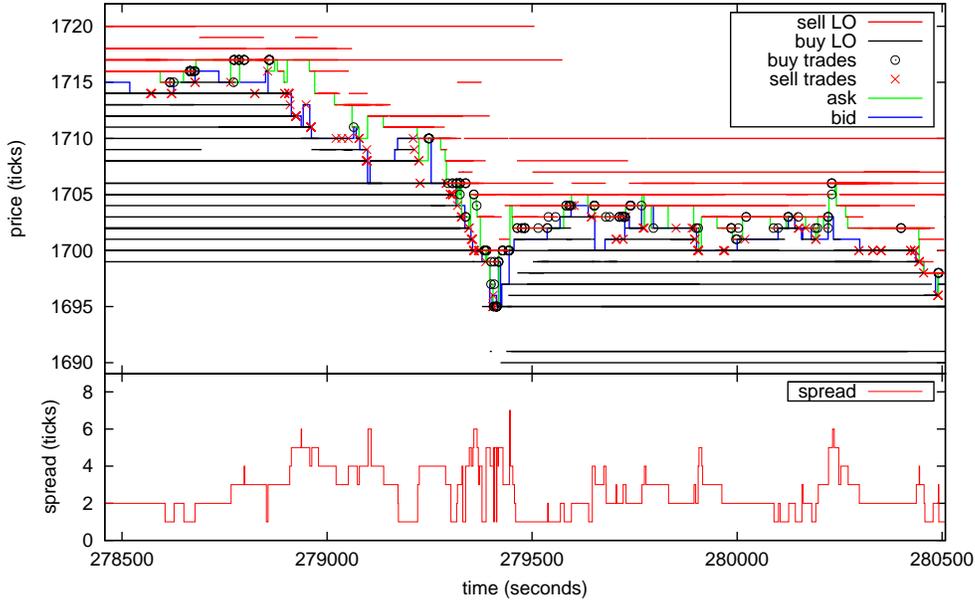}}
\caption{The trading activity of the stock GSK between the $26500-28500$'th seconds of the $8$'th trading day of $2002$. In the top half of the plot price levels with buy/sell limit orders are indicated by black/red horizontal lines. The bid and ask are indicated by blue/green lines. During the period a snapshots of the order book every $50$ seconds. The lack of orders is indicated by black color, buy/sell orders by red/blue, and the bid-ask spread by white. Buyer initiated transactions are marked by black circles, while seller initiated transactions by red crosses.}
\label{fig:ponzi_short8}
\end{figure*}

In order to see more temporal structure one has to move down to the scale of individual events, where the microstructure of the order book truly begins to play a role. Fig. \ref{fig:ponzi_short8} shows a $4000$ second period during the end of the same $8$'th day. To visualize the dataset we used the technique introduced by Ponzi et al. \cite{ponzi.liquidity}. This is the finest level of price formation. Here we will very briefly sketch a qualitative picture as suggested by Refs. \cite{ponzi.liquidity, weber.qf2005, farmer.efficiency, farmer.whatreally}.

During usual trading there is enough volume in a narrow range around the spread to satisfy market orders, and so trading is confined there. The bid/ask can basically change in three ways: (i) A typical market order has a volume which is less or equal to the best opposite limit order. If a market order arrives, it either leaves some volume at the bid/ask so the level does not move, or it removes the entire best offer but usually not more, and so the change in the quote equals the gap between the best and the second best opposite limit price \cite{farmer.whatreally}. (ii) A new limit order is placed inside the spread. (iii) All orders at the best price are canceled. The shift in the bid/ask price due to an event is called its \emph{price impact}. Case (iii) is quite rare (see Sec. VI. of Ref. \cite{ponzi.liquidity}). The immediate effect of cases (i) and (ii) is the opposite, i.e., the spread increases or decreases. The information content of an order placement and an execution is also very different, so it is not surprising that the caused variation in the bid-ask spread will evolve further in different ways. The reaction is faster for an increase in the spread than for a decrease, and the mean reversion is stronger. Moreover, transactions have a stronger effect than other events. Because the type of market orders (buy or sell) are long time correlated, a buy market order for example is likely to induce further buy market orders or to be a reaction to those. Thus even if a buy market order \emph{itself} does not change the ask price, the ask price is likely to increase on average due to subsequent events \cite{ponzi.liquidity}.

Much more rarely one larger or many smaller limit orders arrive, which wipe out a large portion of the book and thus result in an abrupt increase of the bid-ask spread which correspond to a decrease in liquidity. This is followed by intense fluctuations, see for example the period near $t=279500$ sec in Fig. \ref{fig:ponzi_short8}. After the shock the bid-ask spread closes. In our example this takes about $2$ minutes. Complete relaxation of the spread/liquidity can be very slow, but the half-life of the decay is typically of the order $10^2-10^3$ seconds.

The price change corresponding to these large events is still much less than the width of the order book (with some very few exceptions) but they can represent a very substantial fraction of the total volume present at that time, because the orders very far from the best tend to be smaller. After the spread closes typically there is no complete reversion to the price levels before the event, and a new midquote price is formed \cite{ponzi.liquidity, zawa.reaction}.

\section{Conclusions}
\label{sec:conclusions}

In this paper, by using various representations of the order book, we have shown that the qualitative picture that describes the dynamics of trading strongly depends on the time scale. On a monthly scale price changes are well described qualitatively by an ordinary diffusion with a drift, and the fluctuations in supply and demand are not relevant. In contrast, for daily data price changes are very broadly distributed, and the volatility is clustered. This is mirrored by the large day-to-day fluctuations in the order book. The imbalance between the volume of buy and sell limit orders also varies strongly, but this variation is not reflected directly in the returns. The qualitative picture changes dramatically when one moves down to the resolution of individual transactions, where returns are governed by the price impact of individual events and liquidity fluctuations. Despite their potential importance in the execution of high-frequency trading strategies, at present no models can fully account for the rich structure of returns on very short time scales. There has been some recent success along these lines, but we believe that many questions still remain unanswered.

\section*{Acknowledgments} 

This work was supported by COST--STSM--P10--917 and OTKA T049238. JK is member of the Center for Applied Mathematics and Computational Physics, BME. FL acknowledges support from MIUR research project ``Dinamica di altissima frequenza nei mercati finanziari'' and NEST-DYSONET 12911 EU project.

\bibliographystyle{spiebib}
\bibliography{orderbook_firenze2007}

\end{document}